\documentclass{aa}  

\usepackage{xcolor}
\usepackage{txfonts}
\usepackage{graphicx}

\definecolor{darkbrown}{HTML}{8c4600}
\definecolor{darkblue}{HTML}{1833a1}

\newcommand{\vk}{v_{\rm k}}

\newcommand{\vg}{v_{\rm g}}
\newcommand{\vp}{v_{\rm p}}
\newcommand{\vfrag}{v_{\rm frag}}
\newcommand{\vcoll}{v_{\rm coll}}
\newcommand{\vdrift}{v_{\rm drift}}

\newcommand{\Sigmag}{\Sigma_{\rm g}}
\newcommand{\Sigmap}{\Sigma_{\rm p}}

\newcommand{\dt}{\Delta t}

\newcommand{\tgrow}{t_{\rm grow}}
\newcommand{\tdrift}{t_{\rm drift}}

\newcommand{\St}{{\rm St}}
\newcommand{\Stx}{{\rm St}_{\rm x}}
\newcommand{\Sty}{{\rm St}_{\rm y}}

\newcommand{\Stmax}{{\rm St}_{\rm max}}
\newcommand{\Stfrag}{{\rm St}_{\rm frag}}
\newcommand{\Stdrift}{{\rm St}_{\rm drift}}

\newcommand{\cs}{c_{\rm s}}

\newcommand{\Hp}{H_{\rm p}}
\newcommand{\ceff}{\tilde{c}_{\rm s}}
\newcommand{\Heff}{\tilde{H}}
\newcommand{\Omegak}{\Omega_{\rm K}}

\newcommand{\rhop}{\rho_{\rm p}}
\newcommand{\rhog}{\rho_{\rm g}}

\begin{document}

\title{Positive Feedback: How a Synergy Between the Streaming Instability and Dust Coagulation Forms Planetesimals}
\titlerunning{Positive Feedback}

\author{
Daniel Carrera\inst{\ref{NMSU}}\thanks{\email{carrera4@nmsu.edu}}
\and Jeonghoon Lim\inst{\ref{ISU}}
\and Linn E.J. Eriksson\inst{\ref{Stony},\ref{AMNH}}
\and Wladimir Lyra\inst{\ref{NMSU}}
\and Jacob B. Simon\inst{\ref{ISU}}
}

\institute{
New Mexico State University, Department of Astronomy, PO Box 30001 MSC 4500, Las Cruces, NM 88001, USA\label{NMSU}
\and Department of Physics and Astronomy, Iowa State University, Ames, IA 50010, USA\label{ISU}
\and Institute for Advanced Computational Sciences, Stony Brook University, Stony Brook, NY, 11794-5250, USA\label{Stony}
\and Department of Astrophysics, American Museum of Natural History, 200 Central Park West, New York, NY 10024, USA\label{AMNH}
}


\abstract
{One of the most important open questions in planet formation is how dust grains in a protoplanetary disk manage to overcome growth barriers and form the $\sim$100km planet building blocks that we call planetesimals. There appears to be a gap between the largest grains that can be produce by coagulation, and the smallest grains that are needed for the streaming instability (SI) to form planetesimals.}
{Here we explore a novel hypothesis: That dust coagulation and the SI work in tandem. That they form a feedback loop where each one boosts the action of the other to bridge the gap between dust grains and planetesimals.}
{We develop a semi-analytical model of dust concentration due to the SI, and an analytic model of how the SI affects the fragmentation and radial drift barriers. We then combine those to model our proposed feedback loop.}
{In the fragmentation-limited regime, we find a powerful synergy between the SI and dust growth that drastically increases both grain sizes and densities. We find that a midplane dust-to-gas ratio of $\epsilon \ge 0.3$ is a sufficient condition for the feedback loop to reach the planetesimal-forming region for turbulence values $10^{-4} \le \alpha \le 10^{-3}$ and grain sizes $0.01 \le \St \le 0.1$. In contrast, the drift-limited regime only shows grain growth, without significant dust accumulation. Planet formation in the drift-limited portion of the disk may require other processes (particle traps) to halt radial drift.}
{}

\keywords{planetesimals -- planet formation}

\maketitle

\nolinenumbers

\section{Introduction}
\label{sec:intro}

An interesting feature of planet formation is that it is a complex, multi-scale, multi-physics problem. The micro-physics of dust grain dynamics and cohesion, the long-range gravitational torques from the disk, and the intermediate-range physics of turbulence and aerodynamic drag, are all key ingredients that must be studied and understood. By necessity, these processes are generally studied in isolation and we often struggle to understand the interactions between them.

In this work we explore the interaction between two of these processes: Grain growth by coagulation \citep[e.g.,][]{Birnstiel_2012} and the streaming instability \citep[SI,][]{Youdin_2005,Johansen_2007b,Youdin_2007b}. We seek to understand how they combine to form the $\sim100$ km primordial bodies that we call planetesimals, which are thought to be the building blocks of terrestrial planets and giant planet cores \citep{Kokubo_1996,Kokubo_2000}.

\subsection{Streaming instability}
\label{sec:intro:streaming_instability}

The SI \citep{Youdin_2005} is a convergence of radial drift among solid grains. It is perhaps the leading candidate for an explanation of how planetesimals form. While the details are complex, its qualitative behavior is simple enough: As dust grains feel gas drag, they also push back on the gas, and this feedback causes radially-drifting grains to converge into azimuthal filaments. If those filaments reach the Roche density

\begin{equation}\label{eqn:roche_density}
    \rho_{\rm R} = \frac{9\Omegak^2}{4\pi G},
\end{equation}

\noindent
where $\Omegak$ is the Keplerian frequency and $G$ is the gravitational constant, then the particle cloud may collapse gravitationally.

Much of the interest in the SI is driven by evidence that seem to favor it. The size distribution of present-day asteroids and the cratering record of Vesta \citep{Morbidelli_2009}, along with the large number of equal-size Kuiper belt binaries \citep{Nesvorny_2010,Fraser_2017}, suggests that planetesimals formed through the gravitational collapse of a dust cloud. Gravitational collapse is also a neat way of overcoming the growth barriers discussed in the previous section. One piece of evidence that points to the SI in particular is that the SI reproduces the obliquity and angular momentum distributions of trans-Neptunian binaries \citep{Nesvorny_2019,Nesvorny_2021} as well as the densities of Kuiper-Belt Objects \citep{Canas_2024}.

However, the major challenge facing the SI (and a reason to be cautious about it) is that the conditions needed to trigger the SI \citep{Carrera_2015,Yang_2017,Li_2021,Lim_2024a,Lim_2024b} are inconsistent with dust growth and disk evolution models, as it requires fairly large grains and/or high column dust-to-gas ratio $Z$ in order to form filaments dense enough to reach the Roche density. Importantly, most studies of the SI criterion rely on 2D models without external turbulence. \citet{Lim_2024a} showed that even a modest degree of external turbulence can make the critical $Z$ much higher than those turbulence-free 2D models suggest. In this paper we make the case that dust coagulation is the missing ingredient that allows the SI to work with realistic turbulence.

\subsection{Interaction between the SI and dust growth}
\label{sec:intro:boost}

In this paper we explore a simple hypothesis: That the SI and dust growth boost each other, in a positive feedback. The SI leads to enhanced dust concentration, which reduces turbulence and radial drift. This, in turn, boosts grain growth, which then leads to a stronger, more efficient SI. If so, the combination of the SI and dust growth may be the key ingredient that makes planetesimal formation possible. Following a similar thought process, \citet{Tominaga_2023} showed that SI filaments promote dust growth, and \citet{Ho_2024} showed that dust growth can trigger the SI. We posit that there is, in fact, a feedback loop between dust growth and the SI, so that the two processes boost each other far beyond what might appear from studying just one interaction or the other.

This paper is organized as follows: Most of this manuscript consists of the presentation of the model in section \S\ref{sec:model}, where we derive the detailed relationships between the SI, mass loading, turbulence, and radial drift. In section \S\ref{sec:results} we put these ingredients together to obtain our final results. We discuss in section \S\ref{sec:discussion}, and conclude in section \S\ref{sec:conclusion}.

\section{Model Ingredients}
\label{sec:model}

\subsection{How the SI boosts mass loading}
\label{sec:model:mass_loading}

Perhaps the best-known feature of the SI, and the reason it has garnered so much attention, is its ability to collect solid grains into dense filaments. We repurpose the dataset of \citet{Lim_2024a}, who conducted a large number of 3D simulations of the SI with external (i.e. forced) turbulence for $\St \in [0.01, 0.1]$, $Z \in [0.015, 0.4]$, and $\alpha \in [10^{-4}, 10^{-3}]$ where $\St \equiv t_{\rm stop}\Omegak$ is the grain's Stokes number, $Z = \Sigmap/\Sigmag$ is the column dust-to-gas ratio, and $\alpha$ is the squared Mach number of the turbulence. We calculated the typical dust-to-gas ratio $\epsilon$ inside the SI filaments and fit a powerlaw. The details of this procedure are included in Appendix \ref{appendix:mass_loading}, but the final result is that 55-80\% of the particles inside an SI filament will experience a dust-to-gas ratio at least as high as

\begin{equation}\label{eqn:epsilon_fit}
    \epsilon_{\rm SI}
        \approx 1.12
            \left(\frac{Z}{0.01}\right)^{2.65}
            \left(\frac{\St}{0.1}\right)^{1.42}
            \left(\frac{\alpha}{10^{-4}}\right)^{-1.38}.
\end{equation}

\subsection{How mass loading boosts $\Stfrag$}
\label{sec:model:Stfrag}

The approach to mass loading commonly used in the literature is to think of the gas-dust mixture as a single fluid, taking the limit $\St \rightarrow 0$ so that the mixture represents a colloidal suspension. In this limit, one can think of the dust as contributing to the inertia of the fluid but not to its pressure \citep{Chang_2010,Shi_2013,Laibe_2014,Lin_2017,Chen_2018}. This can be expressed as an ``effective'' sound speed

\begin{eqnarray*}
    P &=& \rhog \cs^2 = (\rhog + \rhop) \ceff^2\\
    \ceff &\equiv& \frac{\cs}{\sqrt{1 + \epsilon}}
\end{eqnarray*}

\noindent
where $\ceff$ is the effective sound speed of the colloid.

However, in this work we are explicitly interested in the case where $\St$ may not be negligible, and the fluid might not be well represented by a colloid. Therefore, we take a different approach. Let us posit that turbulence is generated by a finite energy source that has to be partitioned between the gas and dust components

\begin{eqnarray*}\label{eqn:turbulent_energy}
    E &=& \frac{1}{2} \rhog \vg^2
       +  \frac{1}{2} \rhop \vp^2 \\
      &=& \frac{1}{2} \rhog \vg^2
          \left( 1 + \epsilon\frac{\vp^2}{\vg^2} \right)
      = \text{Const}
\end{eqnarray*}

\noindent
Where $\vg$ and $\vp$ are the root-mean-squared velocities of the gas and dust. Several authors \citep{Voelk_1980,Cuzzi_1993,Schraepler_2004} have shown that the random speed of solid grains in turbulence is $\vp = \vg / \sqrt{1 + \St}$. Insert this into Equation \ref{eqn:turbulent_energy}, and treating $\rhog$ as constant we obtain

\begin{equation}
    \vg^2 \left(1 + \frac{\epsilon}{1 + \St}\right)
    = \frac{2E}{\rhog}
    = \text{Const}
\end{equation}

\noindent
Let $v_{\rm g0} \equiv \sqrt{2E/\rhog}$ be the reference gas velocity. This is the gas velocity when $\epsilon \ll 1$, and it roughly equals the initial gas velocity before dust begins to accumulate. Also let $\alpha \equiv v_{\rm g0}^2/\cs^2$.

\begin{equation}\label{eqn:Vgas}
    \vg \sqrt{1 + \frac{\epsilon}{1 + \St}} = v_{\rm g0}
    = \sqrt{\alpha}\cs
\end{equation}

\noindent
Following the literature, we define an effective sound speed $\ceff$ such that $\vg \equiv \sqrt{\alpha}\ceff$

\begin{equation}\label{eqn:ceff}
    \ceff \equiv \cs \sqrt{\frac{1 + \St}{1 + \St + \epsilon}}
\end{equation}

\noindent
Notice that, in the limit as $\St \rightarrow 0$ gives the $\ceff$ for a colloid.

\subsubsection{Fragmentation Limit}

This limit occurs when grain collision speed $\vcoll = \vg\sqrt{3\St}$ reaches the fragmentation speed $\vfrag$. We combine this with Equation \ref{eqn:Vgas} to obtain

\begin{eqnarray}
    \Stfrag
    = \frac{\vfrag^2}{3\vg^2}
    &=& \Stx \left(1 + \frac{\epsilon}{1 + \St} \right)
        \label{eqn:Stfrag_boost}\\
    \text{where}\;\;
    \St_{\rm x} &\equiv& \frac{\vfrag^2}{3\alpha\cs^2}\label{eqn:Stx}
\end{eqnarray}

\noindent
$\Stx$ is the definition of $\Stfrag$ commonly found in the literature. Mass loading boosts the fragmentation barrier by a factor of $1 + \epsilon/(1 + \St)$. Equation \ref{eqn:Stfrag_boost} is a quadratic in $\St$. After simplification, its solution can be written as

\begin{equation}\label{eqn:Stfrag_new}
    \Stfrag = \frac{
        \St_{\rm x}
        - 1 
        + \sqrt{(\St_{\rm x} + 1)^2 + 4\epsilon\St_{\rm x}}
        }{2}
\end{equation}

\noindent
Notice that $\Stfrag = \Stx$ for $\epsilon = 0$ and $\Stfrag > \Stx$ for $\epsilon > 0$. Other limits include $\Stfrag \approx (1 + \epsilon)\Stx$ for $\epsilon < 1$ and $\Stfrag \approx \sqrt{\epsilon\Stx}$ for $\epsilon \gg \Stx + \Stx^{-1}$. But for practical use, the best strategy is to use Equation \ref{eqn:Stfrag_new}.

\subsection{How mass loading boosts $\Stdrift$}
\label{sec:model:drift}

The radial drift barrier occurs when the particle drift rate matches its growth rate. Here, too, the SI can increase $\Stdrift$ as higher $\epsilon$ leads slower radial drift. The steady-state particle drift rate was derived by \citet{Nakagawa_1986}. In the limit where $\St^2 \ll 1$ the drift speed is

\begin{equation}\label{eqn:vdrift_eps}
    \vdrift = \frac{-2\eta\vk}{\St + (1+\epsilon)^2/\St}
\end{equation}

\noindent
For $\epsilon \ll 1$ this matches the formula for $\vdrift$ seen in the literature. But once gain, we want to consider the case where $\epsilon$ is not small. The particle drift timescale is

\begin{equation}\label{eqn:tdrift_eps}
    \tdrift \equiv \frac{r}{\vdrift}
        = \frac{\St + (1+\epsilon)^2/\St}{-2\eta\Omegak}
\end{equation}

\noindent
\citet{Birnstiel_2012} derive the formula for the growth rate $\tgrow$. If we follow their derivation, but omit the final step, we find

\begin{equation}\label{eqn:tgrow}
    \tgrow = \frac{1}{\epsilon\Omegak}
            \sqrt{\frac{\St}{\alpha}}.
\end{equation}

\noindent
\citet{Birnstiel_2012}'s final step is to apply the expression for $\epsilon$ from sedimentation to the midplane. Since we are interested in the $\epsilon$ due to the SI, Equation \ref{eqn:tgrow} is more appropriate for our needs. The radial drift barrier occurs when $\tgrow = \tdrift$. The result is a quartic equation in $\St$.

\begin{equation}
    \St^2 + \frac{2\eta}{\epsilon\sqrt{\alpha}} \St^{3/2} + (1+\epsilon)^2 = 0
\end{equation}

\noindent
This equation has an analytic solution but it is very complicated. To simplify it, we assume that $\St \ll (1+\epsilon)^2/\St$, so that Equation \ref{eqn:tdrift_eps} simplifies to

\begin{equation}
    \tdrift \approx \frac{(1+\epsilon)^2}{-2\eta\Omegak\St}.
\end{equation}

\noindent
With this simplification, we obtain the radial drift limit as

\begin{eqnarray}\label{eqn:Stdrift_new}
    \Stdrift &=& \left[
            \frac{\epsilon^2(1+\epsilon)^4\alpha}{4\eta^2}
        \right]^{1/3}
        = \left[
            \Sty^2
            \frac{\epsilon^2\alpha}{Z^2}
            (1+\epsilon)^4
        \right]^{1/3}\\
    \text{where}
    \;\;\;\;
    \Sty &\equiv&  \frac{-Z}{2\eta}.
\end{eqnarray}

\noindent
$\Sty$ is the definition of $\Stdrift$ commonly found in the literature. To obtain that value, one has to assume $\epsilon \ll 1$ as well as $\epsilon$ from vertical sedimentation.

\subsection{Feedback between dust growth and the SI}
\label{sec:model:feedback}

Finally, we combine all the ingredients. Starting from an initial $(Z,\alpha,\Stx)$ or $(Z,\alpha,\Sty)$, we apply the following algorithm:

\begin{enumerate}
\item Let $t_{\rm SI} = 50\Omegak$ be the filament growth timescale. In truth, $t_{\rm SI}$ depends on $\St$ \citep[e.g.,][]{Carrera_2015,Yang_2017,Li_2021}, but there is no published formula for $t_{\rm SI}$ and $50\Omegak$ is typical.\newline

\item Initialize $\epsilon$ with the midplane dust-to-gas ratio due to vertical sedimentation with particle feedback (section \S\ref{sec:model:feedback:epsilon_init} below).

\item Feedback Loop:

\begin{itemize}
    \item Update $t_{\rm coag}$ using Equation \ref{eqn:tgrow}.
    \item Update $\epsilon_{\rm SI}$ using Equation \ref{eqn:epsilon_fit}.
    \item Let $\dt \coloneqq 0.1\min(t_{\rm coag}, t_{\rm SI})$ be the iteration timestep.
    \item Let $\Stmax \coloneqq \Stfrag$ or $\Stdrift$ (Equations \ref{eqn:Stfrag_new} and \ref{eqn:Stdrift_new}).
    \item Update $(\St,\epsilon)$ at the same time
    \begin{itemize}
        \item $\St \coloneqq \min\left[
                        \Stmax,
                        \St\cdot\exp(\dt/t_{\rm coag})
                    \right]$
        \item $\epsilon \coloneqq \min\left[
                        \epsilon_{\rm SI},
                        \epsilon\cdot\exp(\dt/t_{\rm SI})
                    \right]$
    \end{itemize}
\end{itemize}

\end{enumerate}

\noindent
Treating $\{\Stx,\Sty\}$ as input parameters allows us to explore the problem without assuming one particular disk model. We also examine each growth barrier independently.

\subsubsection{Dust scale height and sedimentation}
\label{sec:model:feedback:epsilon_init}

We start with the $\epsilon$ due to sedimentation and vertical diffusion (i.e. prior to the SI). The usual approach is to compute

\begin{eqnarray}\label{eqn:epsilon_sedimentation}
    \epsilon &=& Z \frac{H}{\Hp}\\
    \text{for}\;\;
    \Hp &=& H\sqrt{\frac{\alpha}{\St + \alpha}}
\end{eqnarray}

\noindent
where $H$ and $\Hp$ are the gas and particle scale heights, and $\rhop$ is assumed to have a vertical Gaussian density distribution, as derived by \citet{Youdin_2007}. However, \citet{Lim_2024a} showed that, due to mass loading, this expression is a poor predictor of $\Hp$, and that the density profile is not Gaussian. They found that

\begin{equation}\label{eqn:Hp}
    \Hp = \frac{H}{\sqrt{1 + \epsilon}}
        \sqrt{
            \left( \frac{\Pi}{5} \right)^2 +
            \frac{\alpha}{\alpha + \St}
        }
\end{equation}

\noindent
is a better predictor of the height of the particle layer. We adopt $\Pi = 0.05$ because that was the value used by \citet{Lim_2024a}. Equation \ref{eqn:Hp} was based on expression for the effective scale height ($\Heff \equiv \ceff/\Omega$) for a colloid \citep{Yang_2020}. Because we only use this formula once, before $(\St,\epsilon)$ have had a chance to grow, it can retain the colloid approximation.

Despite the fact that $\rhop$ might not be Gaussian, Equation \ref{eqn:epsilon_sedimentation} is still the best estimate for $\epsilon$ that we are aware of. Combined with Equation \ref{eqn:Hp} we get a quadratic expression for $\epsilon$ with solution

\begin{eqnarray}\label{eqn:epsilon_init}
    \epsilon_{\rm init} &=&
        \frac{\zeta + \sqrt{\zeta^2 + 4\zeta}}{2}\\
    \text{where}\;\;
    \zeta &\equiv& Z^2
        \left[
            \left( \frac{\Pi}{5} \right)^2 +
            \frac{\alpha}{\alpha + \St}
        \right]^{-1}
\end{eqnarray}

\noindent
There is another root, but it corresponds to $\epsilon < 0$.

\section{Results: Feedback Loop}
\label{sec:results}

\begin{figure*}[t]
\centering
\includegraphics[width=0.49\textwidth]{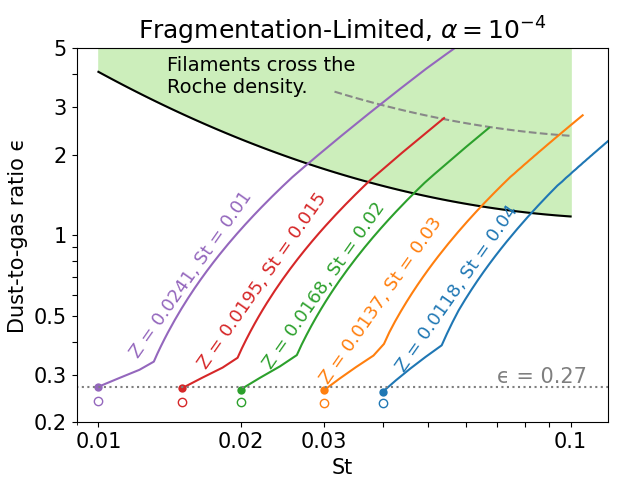}
\includegraphics[width=0.49\textwidth]{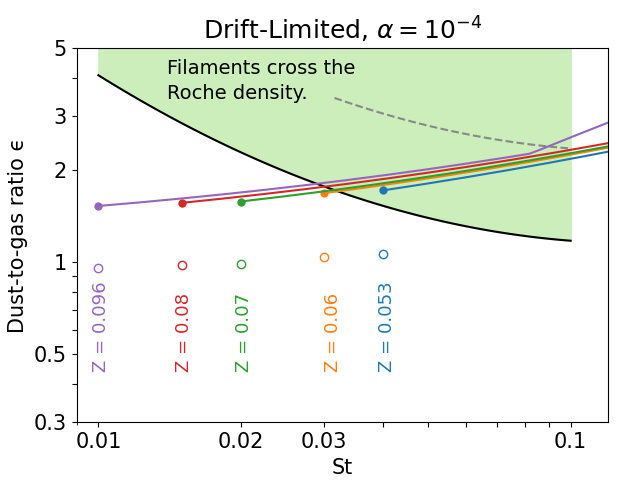}\\
\includegraphics[width=0.49\textwidth]{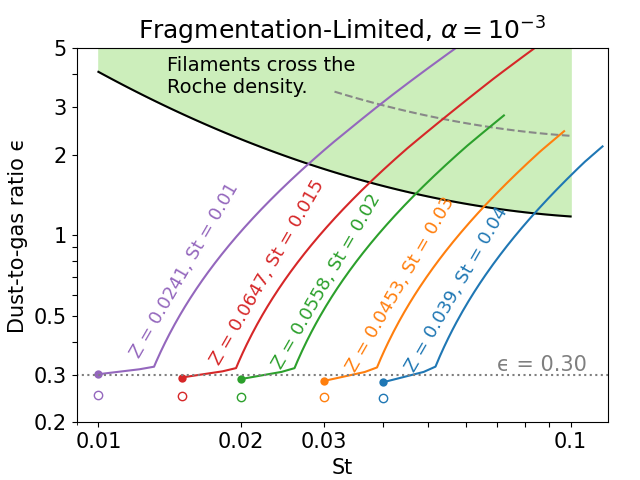}
\includegraphics[width=0.49\textwidth]{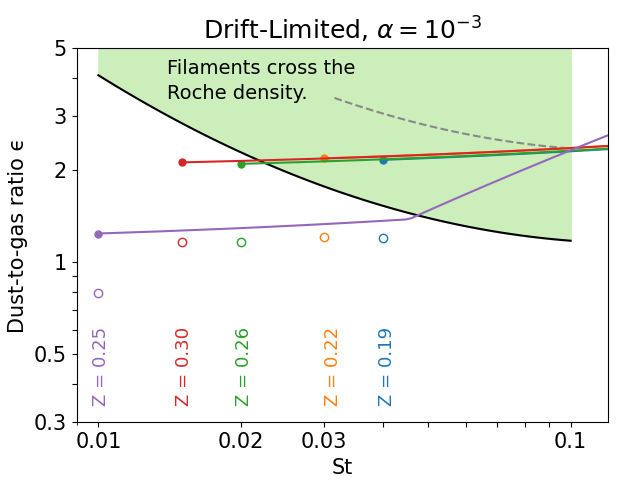}\\
\caption{Particle growth tracks for a range of particle sizes $\St \in [0.01, 0.04]$ in four scenarios: Fragmentation-limited vs drift-limited growth, and $\alpha = 10^{-4}$ or $10^{-3}$. The particle tracks begin at the solid circle, which marks the midplane dust-to-gas ratio for vertical sedimentation in the presence of mass loading (Equation \ref{eqn:epsilon_init}). For reference, we also show an open circle for $\epsilon = Z\sqrt{\St/\alpha}$, which does not account for mass loading. The highlighted region is the range of midplane-averaged $\epsilon$ and $\St$ values where SI filaments cross the Roche density and are likely to experience gravitational collapse. We fine-tuned $Z$ so that the particle tracks would converge at around $2\times$ the SI criterion (dashed line).
\label{fig:particle_tracks}}
\end{figure*}

Figure \ref{fig:particle_tracks} shows our results. We consider either the fragmentation or drift limit, for $\alpha = 10^{-4}$ or $10^{-3}$. For each scenario we compute particle evolution tracks for $0.01 \le \St_{\rm x,y} \le 0.04$ and plot them alongside the SI criterion of \citet{Lim_2024a}. Two important clarifications:

\begin{itemize}
    \item The SI criterion in Figure \ref{fig:particle_tracks} is not a measure of when SI first filaments appear. It is a measure of when those filaments reach the Roche density so they may undergo gravitational collapse.

    \item Our $\epsilon$ is not directly comparable to the SI criterion. In the SI criterion, $\epsilon$ is the average dust-to-gas ratio at the midplane, whereas our $\epsilon$ is the dust-to-gas ratio inside the filaments. For this reason, we also show a dashed line with $2\times$ the SI criterion as a ``margin of safety''.
\end{itemize}

\noindent
For each particle track we show:

\begin{itemize}
    \item An open circle denoting the classic $\epsilon = Z\sqrt{\St/\alpha}$ for the midplane dust-to-gas ratio due to sedimentation.

    \item A closed circle denoting the midplane dust-to-gas ratio when accounting for mass-loading (Equation \ref{eqn:epsilon_init}). This is the start point of the particle track.

    \item The particle track itself, along with a label recording the initial $(Z,\St)$ for that particle track.
\end{itemize}

We find that in every single case, the combination of the SI and coagulation can lead to a drastic increase in particle size. The exact behavior depends primarily on whether the grain sizes are limited by fragmentation or radial drift.

\subsection{Fragmentation Limit}
\label{sec:results:fragmentation}

The SI and coagulation act together to cause a drastic increase in both $\St$ and $\epsilon$. By trial and error we chose the smallest $Z$ that allows each particle track to converge at around $2\times$ the SI criterion. We find that those $Z$ values are quite robust because very small increases in $Z$ lead to much longer tracks. Had we chosen to stop at $1\times$ or $3\times$ the SI criterion, the reported $Z$ values would have been within 2\% of the ones reported here.

We find that, as long as $\epsilon$ starts at $\ge 0.3$, the particle track will go well inside the planetesimal-formation region. If we insert the $\epsilon > 0.3$ criterion into Equation \ref{eqn:epsilon_init}, we obtain

\begin{equation}
    Z > 0.263 \sqrt{\left( \frac{\Pi}{5} \right)^2 + \frac{\alpha}{\alpha + \St}}
\end{equation}

\noindent
For example, for $\alpha = 10^{-4}, \St = 0.01$ we get $Z > 0.026$. But raise $\alpha$ to $10^{-3}$ and the critical $Z$ increases to $Z > 0.078$.

\subsection{Drift Limit}
\label{sec:results:drift}

This case is more subtle: The growth tracks for both barriers exhibit a ``kink''. Initially the growth is mainly in $\St$ with little change in $\epsilon$. At some point, the SI becomes more effective and $\St$ and $\epsilon$ start to grow together. For the fragmentation barrier the kink appears quickly, but for the radial drift barrier the kink occurs so late that it is outside the plotted region for all examples except those that start with $\St = 0.01$. But if our semi-analytical fit for SI clumping extends beyond $\St = 0.1$, that kink could still increase $\epsilon$ in the drift-dominated disk.

\section{Discussion}
\label{sec:discussion}

The origin of planetesimals is a decades-old problem in planet formation. Crossing the vast chasm between mm-cm size grains and 100 km planetesimals is, by far, the largest step needed to form a planet.

We have discovered a powerful feedback loop in which the SI and fragmentation-limited dust coagulation assist each other, to accomplish something that neither one seems able to do alone: form planetesimals under typical disk conditions. That said, it is important to ponder what happens when dust sizes are \textit{not} limited by fragmentation:

\subsection{Drift Barrier}
\label{sec:discussion:drift}

In drift-limited regions there is also a feedback loop which can significantly increase $\St$. However, this is not accompanied by much dust concentration, and seems to require implausible $Z$. We speculate that planet formation in these regions requires dust traps to limit or halt radial drift.

\subsection{Bouncing Barrier}
\label{sec:discussion:bouncing}

Another possibly critical barrier may occur when grain collisions simply do not result in sticking \citep{Zsom_2010}. This barrier is less understood since it depends on grain properties like shape and porosity. If present, it could drastically alter the grain size distribution \citep{Dominik_2024} and lead to grains far smaller than $\Stfrag$, or it might simply be overcome by electrostatic charges \citep{Jungmann_2021}.

With that in mind, it is worth mentioning that the bouncing barrier is similar to fragmentation in that it is a limit in $\vcoll$. Therefore, if present, the bouncing barrier must partake in the same feedback loop that we found for fragmentation. The growth tracks would look like the ones for $\Stfrag$, but probably starting at lower $\St$ and higher $Z$ to reach the planetesimal formation region.

\section{Conclusions}
\label{sec:conclusion}

We have shown that there is a powerful feedback loop in which the SI and dust coagulation boost each other. Each process creates the environment that the other one requires: The SI creates dust-dense regions with low turbulence and slow drift; those are the ideal conditions for dust coagulation to form larger grains, which make the SI more effective. We developed three ingredients:

\begin{enumerate}
\item A semi-analytical estimate of the dust-to-gas ratio $\epsilon$ created by the SI (Equation \ref{eqn:epsilon_fit}).

\item An analytical expression for $\Stfrag$ that models turbulence dampening due to dust feedback (Equation \ref{eqn:Stfrag_new}).

\item An analytical expression for $\Stdrift$ that models the reduced radial drift due to dust feedback (Equation \ref{eqn:Stdrift_new}).
\end{enumerate}

\noindent
In the fragmentation-limited regime, the combination of the SI and dust growth led to a drastic increase in both $\St$ and $\epsilon$, as the two processes fed each other. A practical rule of thumb is that a midplane dust-to-gas ratio of $\epsilon \ge 0.3$ is sufficient to reach the planetesimal-forming region across the full range of $(\St,\alpha)$ that we explored. In other words, the positive feedback between the SI and dust growth seems to resolve a decades-old problem in planet formation, at least in the fragmentation-limited inner disk.

In the drift-limited regime, we saw a dramatic increase in $\St$, but it was not usually accompanied by a significant increase in $\epsilon$, at least for $\St \le 0.1$; but if our semi-analytical fit for SI clumping extends to $\St > 0.1$, it could still lead to significant increase in $\epsilon$ in the drift-dominated disk. Otherwise, planetesimal formation in the outer disk may require dust traps to mitigate radial drift and increase $\epsilon$.

\begin{acknowledgements}
DC, WL, and JBS acknowledge support from NASA under {\em Emerging Worlds} through grant 80NSSC25K7414. J.L. acknowledges support from NASA under the Future Investigators in NASA Earth and Space Science and Technology grant \# 80NSSC22K1322. LE acknowledges the support from NASA via the Emerging Worlds program (\#80NSSC25K7117), as well as funding by the Institute for Advanced Computational Science Postdoctoral Fellowship.
\end{acknowledgements}


\bibliographystyle{aa}
\bibliography{references}


\appendix
\section{How the SI boosts mass loading}
\label{appendix:mass_loading}

Perhaps the best-known feature of the SI, and the reason it has garnered so much attention, is its ability to collect solid grains into dense filaments. Here we repurpose the dataset of \citet{Lim_2024a}, who conducted a large number of 3D simulations of the SI with external (i.e. forced) turbulence. We refer to that paper for details of their simulation setup. In brief, they conducted a systematic exploration of the SI for $\St \in [0.01, 0.1]$, $Z \in [0.015, 0.4]$, and $\alpha \in [10^{-4}, 10^{-3}]$ where $\St \equiv t_{\rm stop}\Omegak$ is the grain's Stokes number, $Z = \Sigmap/\Sigmag$ is the column dust-to-gas ratio, and $\alpha$ is the squared Mach number of the turbulence.

\begin{figure}[t]
\centering
\includegraphics[width=0.49\textwidth]{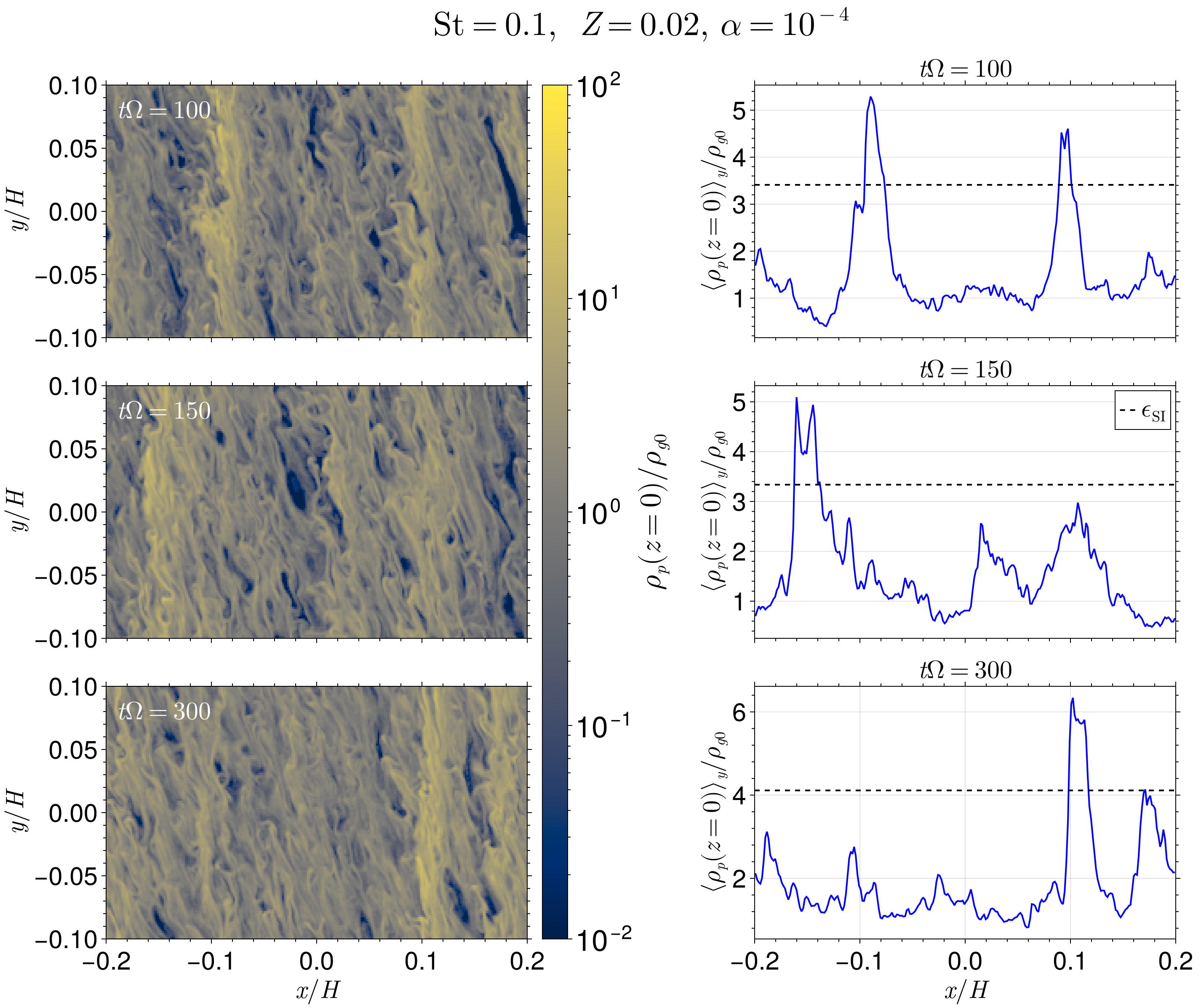}\\
\includegraphics[width=0.49\textwidth]{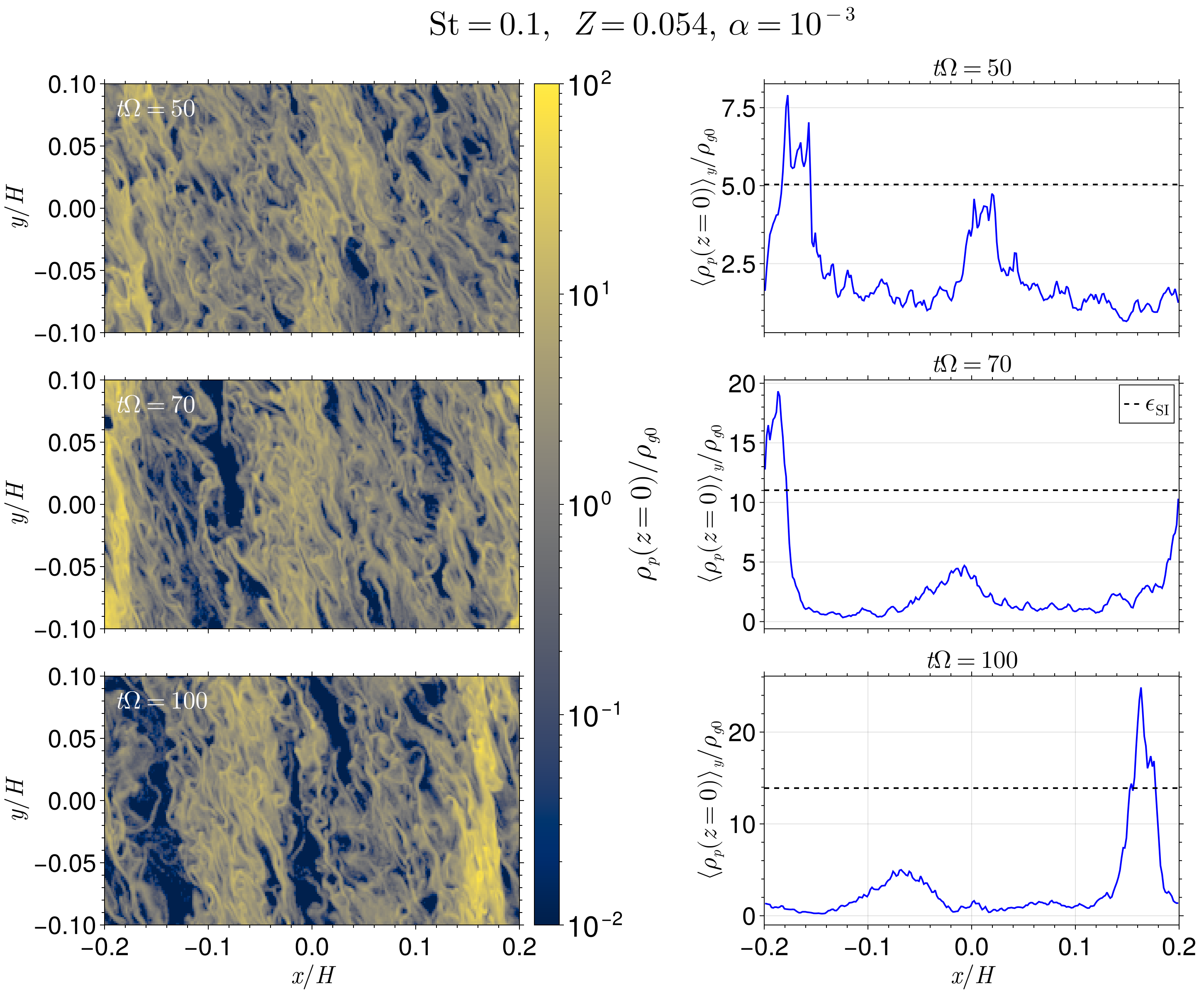}
\caption{Six sample snapshots from two simulations by \citet{Lim_2024a}, showing the midplane particle density ($\rhop(z=0)$, left), and the midplane particle density averaged along the azimuthal direction ($\langle \rhop(z=0) \rangle_y$, right). The value of $\epsilon_{\rm SI}$ is also shown.\label{fig:snapshots}}
\end{figure}

Our goal is to estimate the typical dust-to-gas ratio $\epsilon = \rhop/\rhog$ inside the SI particle filaments. First, we take the mean particle density along the azimuthal direction $\langle \rhop(z=0) \rangle_y$. This removes most of the stochastic noise. Let

\begin{eqnarray}
    \rho_{\rm g,mid} &\equiv& \langle \rhog(z=0) \rangle_{xy}\\
    \rho_{\rm p,mid} &\equiv& \langle \rhop(z=0) \rangle_{xy}\\
    \rho_{\rm p,max} &\equiv& {\rm max}[\langle \rhop(z=0) \rangle_y]\\
    \rho_{\rm p,dev} &\equiv& \rho_{\rm p,max} - \rho_{\rm p,mid}\\
    \epsilon_{\rm SI} &\equiv&
        \frac{1}{\rho_{\rm g,mid}}
        \left(
            \rho_{\rm p,mid} +
            \frac{\rho_{\rm p,dev}}{2}
        \right)
        \label{eqn:epsilon_SI}
\end{eqnarray}

\noindent
where $\rho_{\rm p,dev}$ is the density deviation due to the SI. At ½-deviation, $\epsilon_{\rm SI}$ sits near the middle of the range of $\epsilon$ values inside a filament. Figure \ref{fig:snapshots} shows $\rhop$ and $\epsilon_{\rm SI}$ for six snapshots in two simulations. We have reviewed other snapshots and other simulations, and the results are similar. \textit{Qualitatively,} Equation \ref{eqn:epsilon_SI} appears to be a reliable measure of the typical $\epsilon$ inside an SI filament. \textit{Quantitatively}, we found that most ($55-80$\%) of the particles contained in the filament are in grid cells with $\epsilon > \epsilon_{\rm SI}$. When the simulation first reaches a saturated turbulent state, around $55-70$\% of the particle mass is above $\epsilon_{\rm SI}$, and that value increases to $70-80$\% before the filament reaches the Roche density. That makes $\epsilon_{\rm SI}$ a conservative measure of the density environment experienced by particles inside the filament, and a good choice for exploring dust coagulation inside SI filaments.

\begin{figure}[t]
\centering
\includegraphics[width=0.49\textwidth]{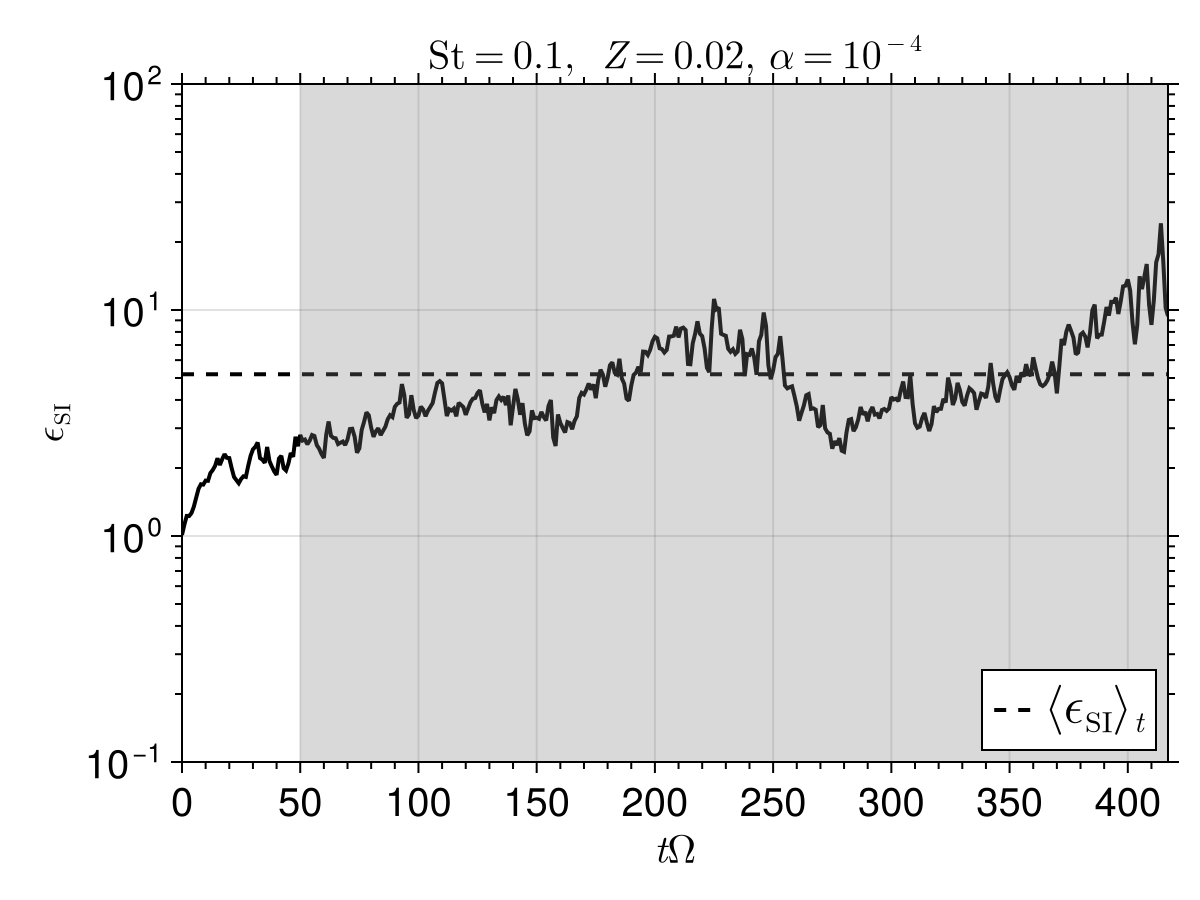}\\
\includegraphics[width=0.49\textwidth]{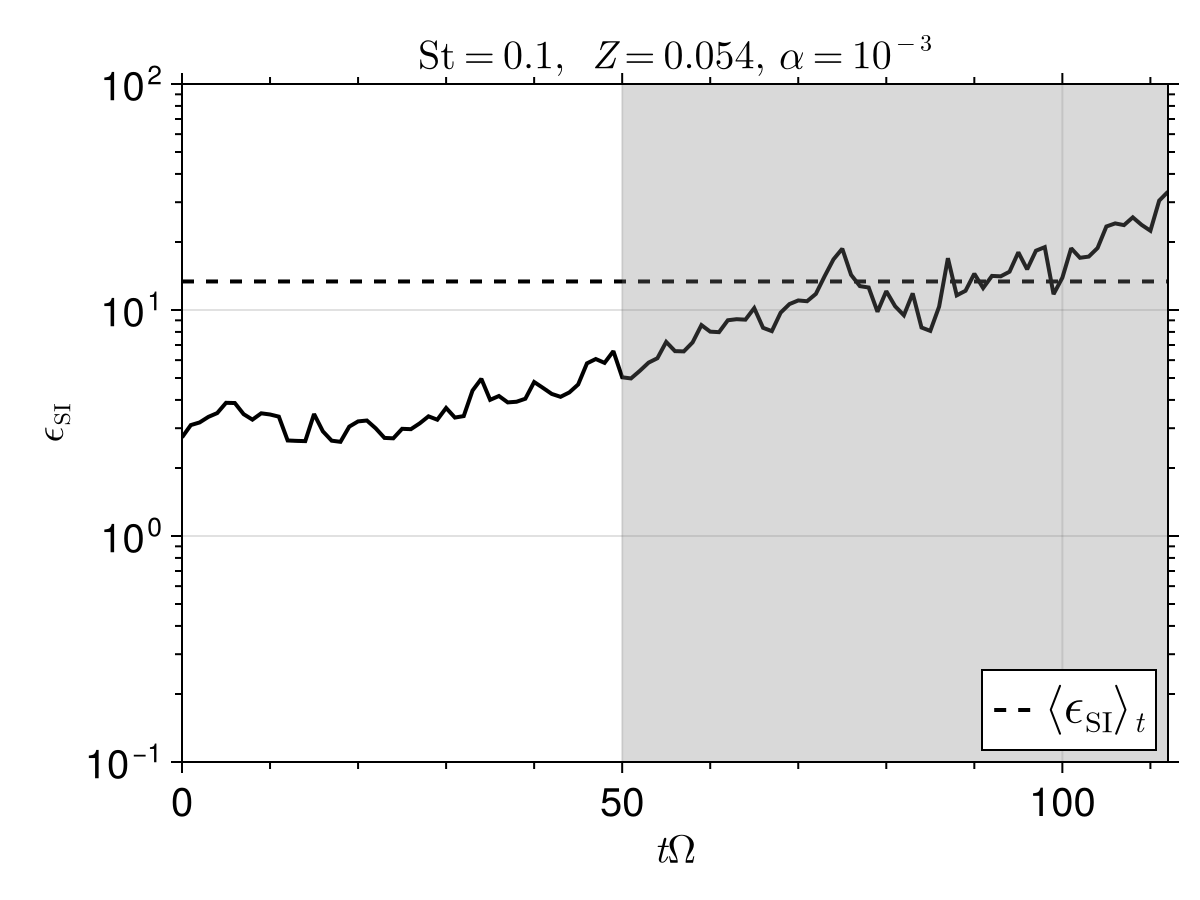}
\caption{Times series of $\epsilon_{\rm SI}$ for two sample runs. The gray region shows the time interval between the moment that the simulation reaches a saturated state, until the moment when $\rhop$ crosses the Roche density (and self-gravity starts to dominate). For each simulation we compute the time average $\langle\epsilon_{\rm SI}\rangle_t$ over this interval.
\label{fig:epsilon_SI_timeseries}}
\end{figure}

For each simulation in \citet{Lim_2024a} we computed $\epsilon_{\rm SI}$ from the moment when $\epsilon_{\rm SI}$ appears to have reached a saturated state, to the moment when $\rhop$ crosses the Roche density. In other words, examine the saturated state prior to gravitational collapse. We compute the time-average $\langle\epsilon_{\rm SI}\rangle_t$ and standard deviation $\sigma_{\epsilon_{\rm SI}}$. Figure \ref{fig:epsilon_SI_timeseries} shows the time series for a sample run, and $\langle \rhop(z=0) \rangle_y$. It also shows the time-averaging interval.

To fit a semi-analytical formula for $\epsilon_{\rm SI}$, we looked for a power-law of the form $\epsilon_{\rm SI} \propto Z^a \St^b \alpha^c$. Writing $\log\epsilon_{\rm SI} = C + a\log Z + b\log\St + c\log\alpha$, we used Ordinary Least Squares to find the fit

\begin{equation}\label{eqn:appendix:epsilon_fit}
    \epsilon_{\rm SI}
        \approx 1.12
            \left(\frac{Z}{0.01}\right)^{2.65}
            \left(\frac{\St}{0.1}\right)^{1.42}
            \left(\frac{\alpha}{10^{-4}}\right)^{-1.38}.
\end{equation}

\noindent
SI filaments exhibit a significantly steeper dependence on all three parameters $(Z,\St,\alpha)$ than the familiar $\epsilon = Z\sqrt{\St/\alpha}$ for dust sedimentation.

\begin{figure}[t]
\centering
\includegraphics[width=0.49\textwidth]{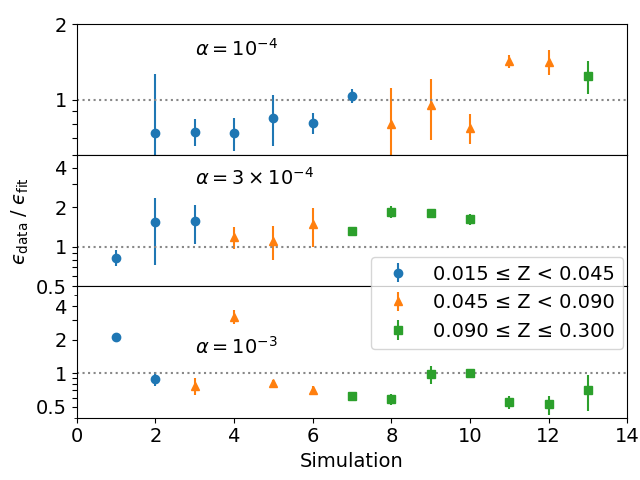}\\
\includegraphics[width=0.49\textwidth]{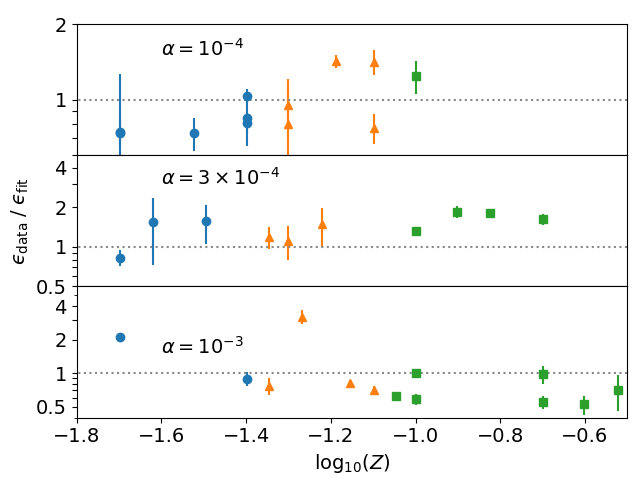}\\
\includegraphics[width=0.49\textwidth]{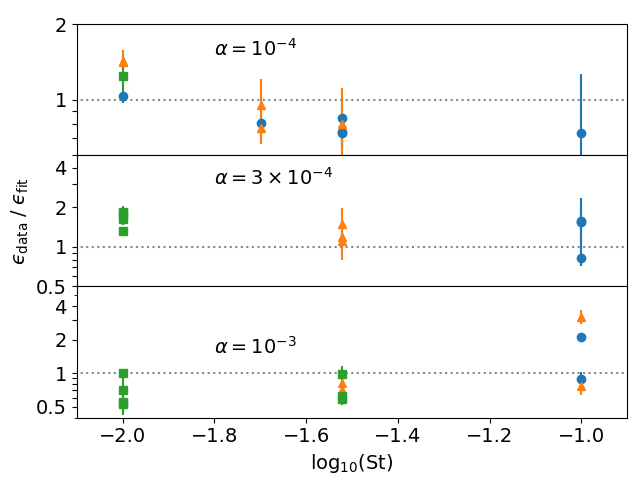}
\caption{We divide our $\{\langle\epsilon_{\rm SI}\rangle_t\}$ dataset across $\alpha$ (top to bottom) and across $Z$ (color coding). We plot the ratio ($\epsilon_{\rm data} / \epsilon_{\rm fit}$) of the dataset versus the fit in Equation \ref{eqn:epsilon_fit}. A perfect fit would mean $\epsilon_{\rm data} / \epsilon_{\rm fit} = 1$. The error bars are $\sigma_{\epsilon_{\rm data}}/\epsilon_{\rm data}$, where $\sigma_{\epsilon_{\rm data}}$ is the standard deviation of the time series.\label{fig:yerror}}
\end{figure}

Figure \ref{fig:yerror} shows the ratio of $\epsilon_{\rm SI}$ from our data, and the best fit in Equation \ref{eqn:appendix:epsilon_fit}. A perfect fit would have $\epsilon_{\rm data} / \epsilon_{\rm fit} = 1$. Notice after applying our fit, there is a slight trend where $\epsilon_{\rm data} / \epsilon_{\rm fit}$ vs $\St$ increases slope with $\alpha$ while $\epsilon_{\rm data} / \epsilon_{\rm fit}$ vs $Z$ decreases slope with $\alpha$. These trends can be removed if we modify the exponents on $Z$ and $\St$ to include components $\propto \log(\alpha)$. However, we judged that the increase in complexity did not justify the small improvement in the fit. For the rest of this work we will use Equation \ref{eqn:epsilon_fit}. Lastly, the error bars in Figure \ref{fig:yerror} are equal to $\sigma_{\epsilon_{\rm SI}}/\langle\epsilon_{\rm SI}\rangle_t$.

\end{document}